\documentclass[iop]{emulateapj}
\usepackage{natbib}
\usepackage{color}
\usepackage[english]{babel}
\usepackage[normalem]{ulem}
\usepackage{blindtext}
\usepackage{textgreek}
\newcommand{\etal}{et\,al.}
\newcommand{\halpha}{H$\alpha$}
\newcommand{\HI}{H~{\sc I}} 
\newcommand{\lsim}{\raise0.3ex\hbox{$<$}\kern-0.75em{\lower0.65ex\hbox{$\sim$}}}
\newcommand{\gsim}{\raise0.3ex\hbox{$>$}\kern-0.75em{\lower0.65ex\hbox{$\sim$}}}
\newcommand{\msun}{M$_{\odot}$}

\newcommand{\kms}{km\,s$^{-1}$}
\begin{document}
\slugcomment{Accepted for publication in the Astronomical Journal}
%-----------------------------------------------------------------------------%
%\title{HI Spectral Line Observations of the Nearby, Star-Forming Dwarf Galaxy NGC 5238}
\title{Rotational Dynamics and Star Formation in the Nearby Dwarf Galaxy NGC 5238}
%-----------------------------------------------------------------------------%

\author{
John M. Cannon\altaffilmark{1}, 
Andrew T. McNichols\altaffilmark{1,2},
Yaron G. Teich\altaffilmark{1,3},
Catherine Ball\altaffilmark{1}, 
John Banovetz\altaffilmark{1,4,5}, 
Annika Brock\altaffilmark{1},
Brian A. Eisner\altaffilmark{1},
Kathleen Fitzgibbon\altaffilmark{1},
Masao Miazzo\altaffilmark{1},
Asra Nizami\altaffilmark{1}, 
Bridget Reilly\altaffilmark{1}, 
Elizabeth Ruvolo\altaffilmark{1},
Quinton Singer\altaffilmark{1}}

\altaffiltext{1}{Department of Physics \& Astronomy, Macalester College, 1600
  Grand Avenue, Saint Paul, MN 55105, USA}
\altaffiltext{2}{NRAO Charlottesville, 520 Edgemont Road, Charlottesville, VA 
22903-2475, USA}
\altaffiltext{3}{Science Department, Somerville High School, 81 Highland Avenue,
Somerville, MA 02143}
\altaffiltext{4}{Department of Physics, MS-B1807, Hamline University,
1536 Hewitt Avenue, Saint Paul, MN 55104, USA}
\altaffiltext{5}{Department of Physics and Astronomy, Purdue University, 
525 Northwestern Avenue, West Lafayette, IN 47907, USA}

%-----------------------------------------------------------------------------%
\begin{abstract}
%-----------------------------------------------------------------------------%

We present new \HI\ spectral line images of the nearby low-mass galaxy
NGC\,5238, acquired with the {Karl G. Jansky Very Large Array
  (VLA}\footnote{The National Radio Astronomy Observatory is a
  facility of the National Science Foundation operated under
  cooperative agreement by Associated Universities, Inc.}).  Located
at a distance of 4.51\,$\pm$\,0.04 Mpc, NGC\,5238 is an actively
star-forming galaxy with widespread \halpha\ and UV continuum
emission.  The source is included in many ongoing and recent nearby
galaxy surveys, but until this work the spatially resolved qualities
of its neutral interstellar medium have remained unstudied.  Our
\HI\ images resolve the disk on physical scales of $\sim$400 pc,
allowing us to undertake a detailed comparative study of the gaseous
and stellar components.  The \HI\ disk is asymmetric in the outer
regions, and the areas of high \HI\ mass surface density display a
crescent-shaped morphology that is slightly offset from the center of
the stellar populations.  The \HI\ column density exceeds 10$^{21}$
cm$^{-2}$ in much of the disk.  We quantify the degree of
co-spatiality of dense \HI\ gas and sites of ongoing star formation as
traced by far-UV and \halpha\ emission.  The neutral gas kinematics
are complex; using a spatially-resolved position-velocity analysis, we
infer a rotational velocity of 31\,$\pm$\,5 \kms.  We place NGC\,5238
on the baryonic Tully-Fisher relation and contextualize the system
amongst other low-mass galaxies.

\end{abstract}						

\keywords{galaxies: evolution --- galaxies: dwarf --- galaxies:
irregular --- galaxies: individual (NGC\,5238)}

%-----------------------------------------------------------------------------%
\section{Introduction}
\label{S1}
%-----------------------------------------------------------------------------%

Nearby, star-forming dwarf galaxies are important laboratories for
probing our understanding of the internal processes that govern
low-mass halos.  Systems within $\sim$10 Mpc can be resolved into
individual stars, allowing detailed study of the interplay between
recently-formed massive stars and the surrounding interstellar medium
(ISM).  With multi-wavelength supporting data and sufficient angular
resolution of the \HI\ 21-cm spectral line, we can also probe the
dynamics of the sources and constrain the ratios of dark to baryonic
matter.

The dwarf galaxies in the local volume have received significant
attention in recent major surveys that span the electromagnetic
spectrum.  An incomplete list includes the GALEX ultraviolet (UV)
imaging survey by \citet{lee11}, the Hubble Space Telescope (HST)
LEGUS UV survey by \citet{calzetti15}, the HST optical survey ANGST by
\citet{dalcanton09}, and the SINGS, LVL, and KINGFISH surveys in the
infrared \citep{kennicutt03,dale09,kennicutt11}.  Taken as a whole,
these surveys have revolutionized our understanding of low-mass
galaxies; the panchromatic approach touches on multiple and diverse
aspects of galaxy evolution, from massive star formation to feedback
to radiation balance in the ISM.

From the \HI\ perspective, the nearby galaxy population has been
studied in significant detail in multiple major interferometric
surveys: WHISP (Westerbork \ion{H}{1} Survey of Irregular and Spiral
Galaxies; {Swaters 2002}\nocite{swaters2002}), FIGGS (Faint Irregular
Galaxies GMRT Survey; {Begum \etal\ 2008}\nocite{begum2008}), SHIELD
(the Survey of \HI\ in Extremely Low-mass Dwarfs; {Cannon
  \etal\ 2011}\nocite{cannon11b}, {Teich \etal\ 2016}\nocite{teich16},
{McNichols \etal\ 2016}\nocite{mcnichols16}), VLA-ANGST (Very Large
Array Survey of ACS Nearby Galaxy Survey Treasury Galaxies; {Ott
  \etal\ 2012}\nocite{ott2012}), LITTLE-THINGS (Local Irregulars That
Trace Luminosity Extremes in The \ion{H}{1} Nearby Galaxy Survey;
{Hunter \etal\ 2012}\nocite{hunter2012}), and LVHIS (the Local Volume
\ion{H}{1} Survey; {Kirby 2012}\nocite{kirby2012}).  These surveys
have provided a nearly complete observational census of the neutral
hydrogen properties of the star-forming dwarfs in the local universe.

Since each of the aforementioned \HI\ surveys has different selection
criteria, there exist multiple local volume dwarf irregular galaxies
that have not yet been studied in detail in the \HI\ spectral line.
Some of these systems also have extensive multi-wavelength
observations from the UV to the infrared (e.g., from the
aforementioned nearby galaxy surveys).  One such system is NGC\,5238
(also known as UGC\,8565, VV\,828, Mrk\,1479, SBS\,1332$+$518,
I\,Zw\,64).  Originally cataloged in \citet{dreyer1888}, this nearby,
actively star-forming galaxy has appeared in more than 100
peer-reviewed manuscripts.  Remarkably, interferometric measurements
of its neutral hydrogen have not been published until the present
work.  The proximity, low mass, and active star formation make it an 
interesting target for detailed observations of the neutral ISM.

NGC\,5238 is resolved into stars in Sloan Digitized Sky Survey (SDSS)
images, which reveal a blue stellar population and numerous prominent
stellar clusters; the central cluster shows very strong nebular
emission lines.  Broad-band HST images are used to derive a distance
of 4.51\,$\pm$\,0.04 Mpc in \citet{tully09}; we adopt this distance
throughout the present work, and show the resulting HST images below.
NGC\,5238 is included in the aforementioned LVL \citep{dale09}, GALEX
UV \citep{lee11}, and LEGUS \citep{calzetti15} surveys.
\citet{moustakas06} use strong-line methods to derive a gas-phase
oxygen abundance 12\,$+$\,log(O/H) $=$ 7.96\,$\pm$0.20 (see also
{Marble \etal\ 2010}\nocite{marble10}), corresponding to Z $\simeq$
19\%\,Z$_{\odot}$ using the Solar oxygen abundance of
\citet{asplund09}.  Modifying the relevant values derived in those
works for the updated distance, we collect global physical properties
of NGC\,5238 in Table~1.

\begin{deluxetable}{lcc}  
\tablecaption{Basic Characteristics of NGC\,5238} 
\tablewidth{0pt}  
\tablehead{ 
\colhead{Parameter} &\colhead{Value}}    
\startdata      
Right ascension (J2000)          &13$^{\rm h}$ 34$^{\rm m}$ 42.$^{\rm s}$5\\        
Declination (J2000)              &+51\arcdeg 36\arcmin 49\arcsec\\    
Adopted distance (Mpc)           &4.51\,$\pm$\,0.04\tablenotemark{a}\\
M$_{\rm B}$ (Mag.)                &$-$14.61\tablenotemark{b}\\
M$_{\star}$ (\msun)               &8.9\,$\times$\,10$^{7}$\tablenotemark{c}\\
Single-dish S$_{\rm HI}$ (Jy km\,s$^{-1}$)          &5.77\,$\pm$\,0.61\tablenotemark{d}\\
Interferometric S$_{\rm HI}$ (Jy km\,s$^{-1}$)           &4.56\,$\pm$\,0.46\tablenotemark{e}\\
Single-dish \HI\ mass M$_{\rm HI}$ (\msun)   &(2.8$\pm$\,0.3)\,$\times$\,10$^7$\tablenotemark{d}\\
Interferometric \HI\ mass M$_{\rm HI}$ (\msun)   &(2.2$\pm$\,0.3)\,$\times$\,10$^7$\tablenotemark{e}\\
\enddata     
\label{table1}
\begin{small}
\tablenotetext{a}{\citet{tully09}} 
\tablenotetext{b}{Modifying M$_{\rm B}$ from {Cook \etal\ (2014a)}\nocite{cook14a} for 
the adopted distance.}
\tablenotetext{c}{Modifying M$_{\star}$ from {Cook \etal\ (2014b)}\nocite{cook14b} 
for the adopted distance.}
\tablenotetext{d}{\citet{thuan99}; note that those authors apply a
  $+$16\% increase in the observed flux integral 
  (4.98\,$\pm$\,0.53 Jy km\,s$^{-1}$) for beam effects.}
\tablenotetext{e}{This work.}
\end{small}
\end{deluxetable}   

The neutral hydrogen properties of NGC\,5238 have only been studied
with single-dish observations.  Using the Nan\c{c}ay 300-m telescope,
\citet{thuan99} find W$_{\rm 50}$ = 32\,$\pm$\,4 \kms\ and an observed
\HI\ line integral S$_{\rm HI}$ $=$ 4.98\,$\pm$\,0.53
Jy\,\kms\ (corrected to 5.77\,$\pm$\,0.61 Jy\,\kms\ for beam effects).
At our adopted distance, NGC\,5238 is a relatively low \HI\ mass
system, with M$_{\rm HI}$ $=$ (2.8$\pm$0.3)\,$\times$\,10$^7$ \msun;
including a 35\% correction for Helium and other metals, the total gas
mass of NGC\,5238 is (3.7$\pm$0.3)\,$\times$\,10$^{7}$ \msun.  The
source is comparable to the most massive galaxies in the SHIELD sample
\citep{teich16,mcnichols16}; that program explicitly studies systems
that inhabit the important but previously understudied mass range
log(M$_{\rm HI}$) $\lsim$ 10$^{7.2}$, and we refer the reader to
\citet{teich16} and \citet{mcnichols16} for details.

We organize this paper as follows.  \S~\ref{S2} presents details about
the new \HI\ observations and the supporting datasets.  \S~\ref{S3}
presents the \HI\ properties of NGC\,5238, including a quantitative
comparison of the degree of co-spatiality between star formation rate
tracers (\halpha\ and far-UV emission) and \HI\ mass surface
densities.  \S~\ref{S4} examines the complex neutral gas kinematics of
NGC\,5238.  In \S~\ref{S5} we contextualize NGC\,5238 against the SHIELD
galaxies, and in \S~\ref{S6} we draw our conclusions.

%-----------------------------------------------------------------------------%
\section{Observations and Data Handling}
\label{S2}
\subsection{HI Observations}
\label{S2.1}
%-----------------------------------------------------------------------------%

HI spectral-line observations of NGC\,5238 were acquired with the Karl
G. Jansky Very Large Array (VLA) in the C configuration on February 1,
2016 for program TDEM\,0021.  This program was executed under the
auspices of the ``Observing for University Classes'' program, a
service provided by the NRAO as an opportunity for courses teaching
radio astronomy theory to acquire new observations to be analyzed by
students.  All of the results in this manuscript stem from the
efforts of undergraduate students at Macalester College.  

The WIDAR correlator divides a 4.0 MHz bandwidth into 1,024
channels, delivering a native spectral resolution of 0.86
km\,s$^{-1}$\,ch$^{-1}$.  The primary and phase calibrators were
J1331+305 (3C\,286) and J1400$+$621, respectively.  The total
on-source integration time was approximately 1.6 hours.  The VLA data
were calibrated using standard prescriptions in the
AIPS\footnote{Developed and maintained by NRAO} and
CASA\footnote{https://casa.nrao.edu} environments.  Imaging of the
J1400+6210 phase calibrator field yielded a flux density S$_{\nu}$ $=$
4.24\,$\pm$\,0.01 Jy.  Continuum subtraction of the NGC\,5238 field
was performed in the $uv$ plane using a first-order fit to line-free
channels bracketing the galaxy in the central 50\% of the bandpass.

Imaging of the calibrated $uv$ visibilties followed standard
prescriptions similar to those described in \citet{cannon15} and in
\citet{teich16}.  In brief, the input continuum-subtracted database
was spectrally smoothed by a factor of two, to produce a velocity
resolution of 7.812 kHz\,ch$^{-1}$ (1.56 km\,s$^{-1}$\,ch$^{-1}$).
The cube was then inverted and cleaned using the IMAGR task in AIPS;
the ROBUST factor is 0.5.  Cleaning was performed to 2.5 times the rms
noise per channel in line-free channels.  Residual flux rescaling was
enforced during the imaging process \citep{jorsater95}.  The resulting
synthesized beam size of 17.69\arcsec\,$\times$\,15.40\arcsec\ was
smoothed to a circular 18\arcsec\ beam.  The rms noise in the final
18\arcsec\ cube is 1.4 mJy\,Bm$^{-1}$.

Two-dimensional moment maps were produced from the three-dimensional
datacube as follows.  The original
17.69\arcsec\,$\times$\,15.40\arcsec\ cube was first spatially
smoothed by a Gaussian kernel to a beam size of 30\arcsec.  Using the
rms noise in this cube, a threshold mask was applied at the
2.5\,$\sigma$ level.  The resulting cube was then examined by hand to
isolate real emission, which is required to be both spectrally and
spatially coincident across three neighboring channels.  The resulting
blanked cube was used as a transfer mask against the 18\arcsec\ beam
cube.  The result was collapsed to create traditional moment maps
representing \HI\ mass surface density, intensity weighted velocity
field, and velocity dispersion.

\begin{figure*}
\epsscale{1.2}
\plotone{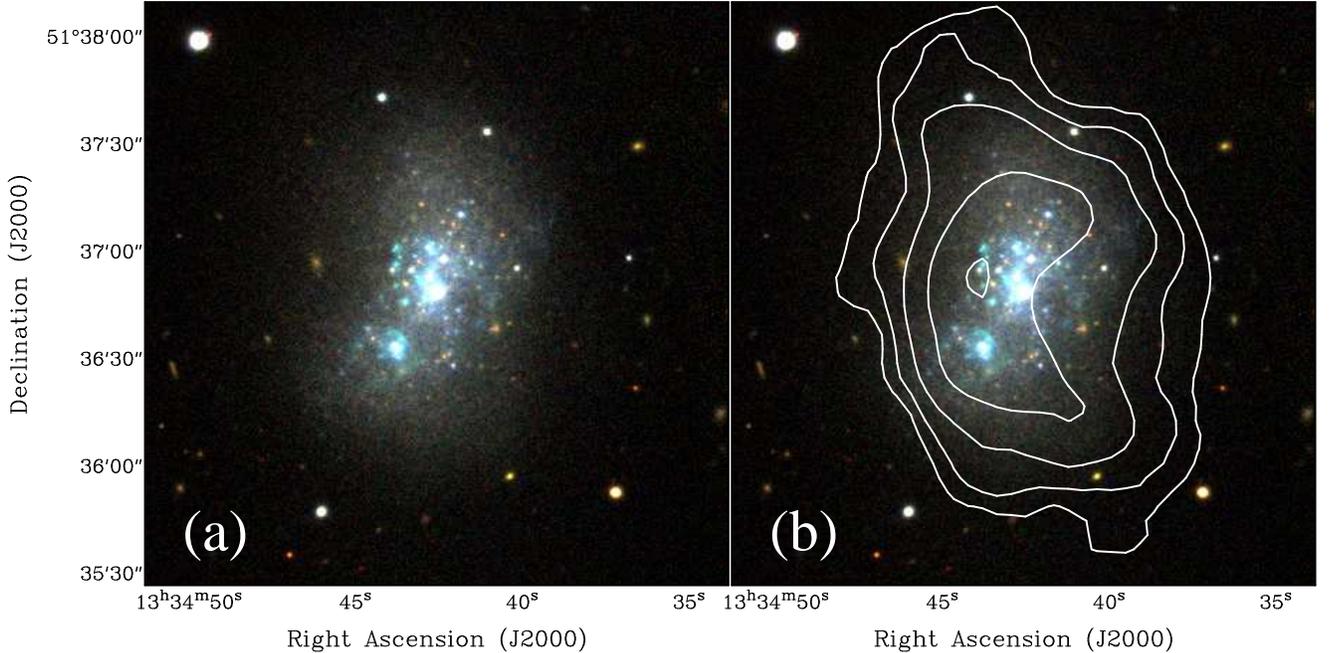}
\epsscale{1.0}
\caption{SDSS 3-color image of NGC\,5238; SDSS g,r,i filters are blue,
  green, and red, respectively.  NGC\,5238 is marginally resolved into
  individual stars in ground-based images.}
\label{figcap1} 
\end{figure*}

%-----------------------------------------------------------------------------%
\subsection{Multiwavelength Archival Observations}
\label{S2.2}
%-----------------------------------------------------------------------------%

We compare our new \HI\ images with archival ground and space-based
imaging in order to study the stellar populations in NGC\,5238.
Calibrated images in the SDSS $g$, $r$, and $i$ filters were obtained
from the SDSS public website \citep{york00}.  Archival HST images from
programs 10905 (Advanced Camera for Surveys, F606W and F814W filters)
and 13364 (Wide Field Planetary Camera 3, F275W, F336W, F438W filters,
from the LEGUS survey; {Calzetti et al. 2015}\nocite{calzetti15})
allow us to study the stellar populations at the highest angular
resolutions, from the near-UV to the near-IR.  We use these images to
construct color mosaic images of NGC\,5238.  The LEGUS data
\citep{calzetti15} highlight the young stellar populations, while the
optical data highlight the older stellar populations (and some nebular
emission).

Continuum-subtracted \halpha\ images are acquired from the LVL survey
data products \citep{kennicutt08}.  The Bok 2.3\,m telescope was used
to image NGC\,5238 in the standard \halpha\ and R-band filters. A
bright foreground star is located in close angular proximity to the
central stellar cluster in NGC\,5238.  This object leaves artifacts
from continuum subtraction in the field; we discuss this source in
more detail below.  The total \halpha\ luminosity of NGC\,5238 derived
by \citet{kennicutt08} is corrected for foreground extinction.
Scaling this value from 5.2 Mpc to the new distance of 4.51 Mpc, we
find log(L$_{\rm H\alpha}$) = 39.09, corresponding to log(SFR$_{\rm
  H\alpha}$) = $-$2.01 M$_{\odot}$\,yr$^{-1}$ using the star formation
rate calibration from \citet{kennicutt98}.

GALEX far-UV (FUV) and near-UV (NUV) imaging of NGC\,5238 is presented
in \citet{lee11}.  The source is detected at high significance in both
filters, indicating recent star formation over the last few hundred
Myr. \citet{lee11} calculates the total extinction-corrected FUV
magnitude from NGC\,5238, m$_{\rm FUV}$ = 15.19\,$\pm$\,0.05.  Using
the star formation rate calibration derived in \citet{mcquinn15}, this
corresponds to log(SFR$_{\rm FUV}$) = $-$1.81 M$_{\odot}$\,yr$^{-1}$
at our adopted distance; note that this star formation rate is
slightly higher than using the star formation rate metric from
\citet{hao11}, which gives log(SFR$_{\rm FUV}$) = $-$2.00
M$_{\odot}$\,yr$^{-1}$.

%-----------------------------------------------------------------------------%
\section{Neutral Gas and Star Formation in NGC\,5238}
\label{S3}
\subsection{The HI Properties of NGC\,5238}
\label{S3.1}
%-----------------------------------------------------------------------------%

In Figure~\ref{figcap1} we present a first-order comparison of the
gaseous and stellar components of NGC\,5238.  The \HI\ mass surface
density distribution (discussed in more detail below) is overlaid as
contours on a 3-color SDSS image.  The outer \HI\ disk is asymmetric;
the morphological major axis is elongated from northeast to southwest.
The high column density \HI\ gas shows a crescent-shaped morphology.
The \HI\ surface density maximum is not coincident with the central
optical peak, but rather is offset to the northeast by $\sim$300 pc;
such offsets are not unusual in dwarf galaxies (see, e.g., {van~Zee
  \etal\ 1997}\nocite{vanzee97}; {Hunter
  \etal\ 1998}\nocite{hunter98}).  The \HI\ disk is slighlty larger
than the high surface brightness optical body; the SDSS Petrosian-r
radius (45\arcsec) can be compared to the size of the \HI\ disk
($\sim$90\arcsec\ radius at the level of 10$^{20}$ cm$^{-2}$).

\begin{figure*}
\epsscale{1.2}
\plotone{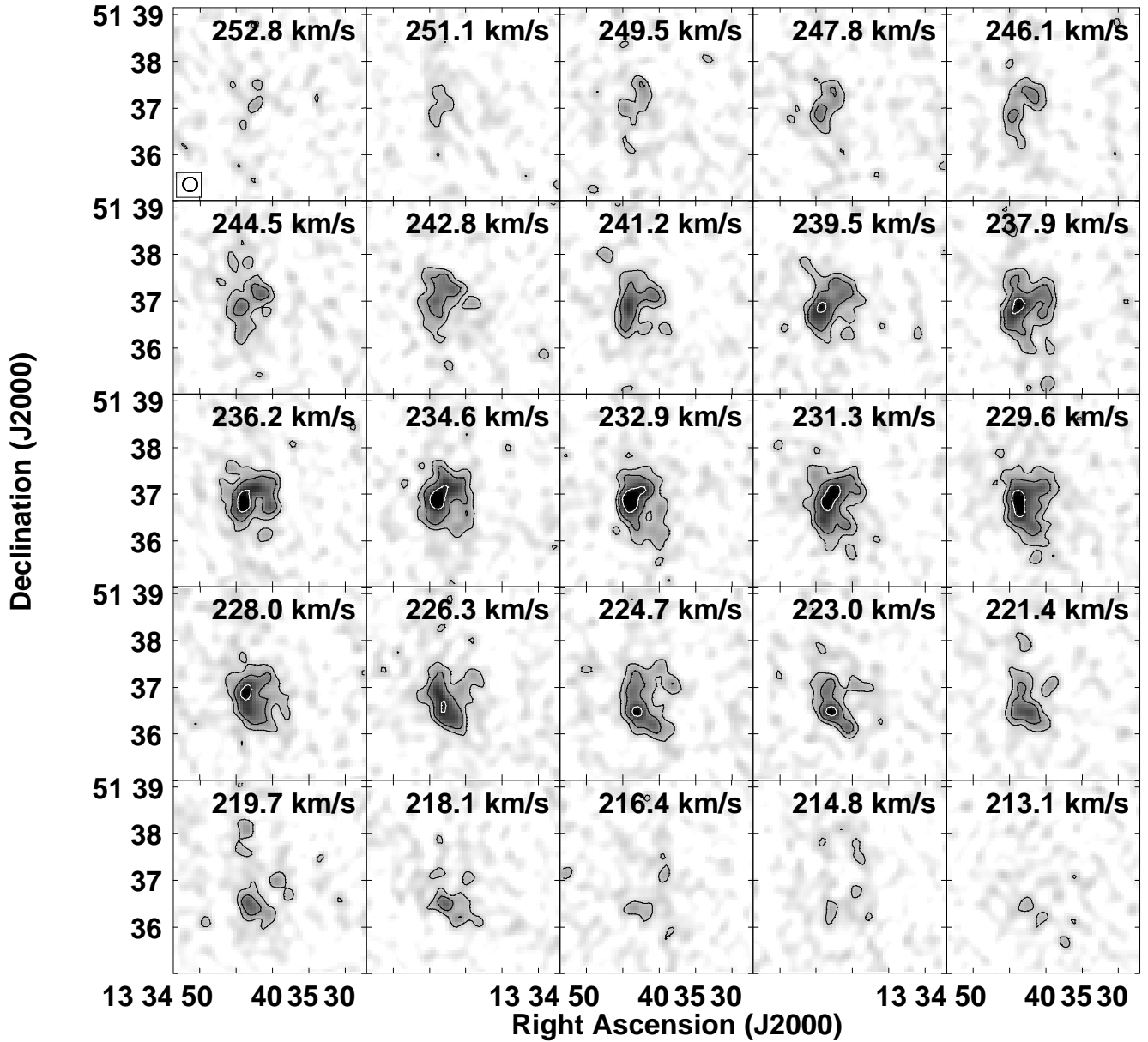}
\epsscale{1.0}
\caption{Channel maps of NGC\,5238.  Contours are overlaid at
  3$\sigma$, 6$\sigma$, and 12$\sigma$, where $\sigma$ is the rms
  noise in line-free channels of the HI cube ($\sigma$ =
  1.4\,$\times$\,10$^{-3}$ Jy\,Bm$^{-1}$).  The circular
  18\arcsec\ beam size is shown in the bottom left of the first
  panel.}
\label{figcap2}
\end{figure*}

\begin{figure*}
\epsscale{1.0}
\plotone{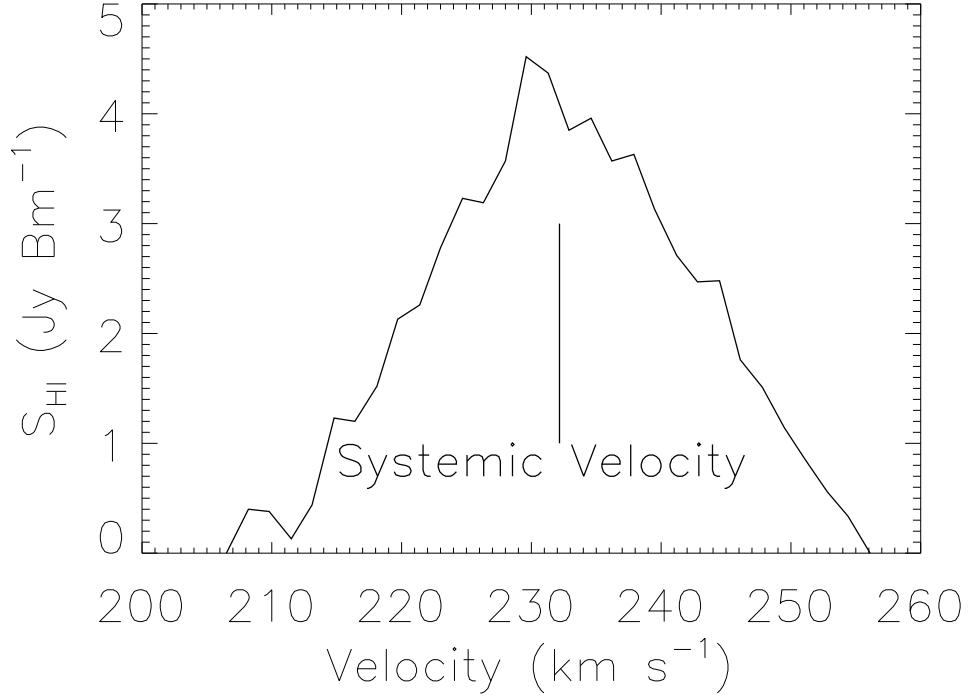}
\caption{Global \HI\ profile of NGC\,5238, created by summing the flux
  in each channel of the 18\arcsec\ resolution blanked datacube. The \HI\
  systemic velocity, derived from an intensity weighted mean of the
  spectrum, is V$_{\rm HI}$ 232\,$\pm$\,1 \kms.}
\label{figcap3}
\end{figure*}

The \HI\ spectral line is detected at high significance across roughly
25 channels ($\sim$40 \kms) of the final 18\arcsec\ datacube.
Figure~\ref{figcap2} shows individual channel maps of the
\HI\ emission across which gas is detected.  Summing the emission
across these channels produces the global \HI\ profile shown in
Figure~\ref{figcap3}.  As expected based on the galaxy's low mass, the
profile is Gaussian-like with no obvious double-horn structure.  Using
the intensity-weighted average of the global \HI\ profile, the 
systemic velocity is V$_{\rm HI}$ $=$ 232\,$\pm$\,1 \kms; this is in
good agreement with V$_{\rm HI}$ 235\,$\pm$\,2 \kms\ derived in
\citet{thuan99}.

\begin{figure*}
\epsscale{1.2}
\plotone{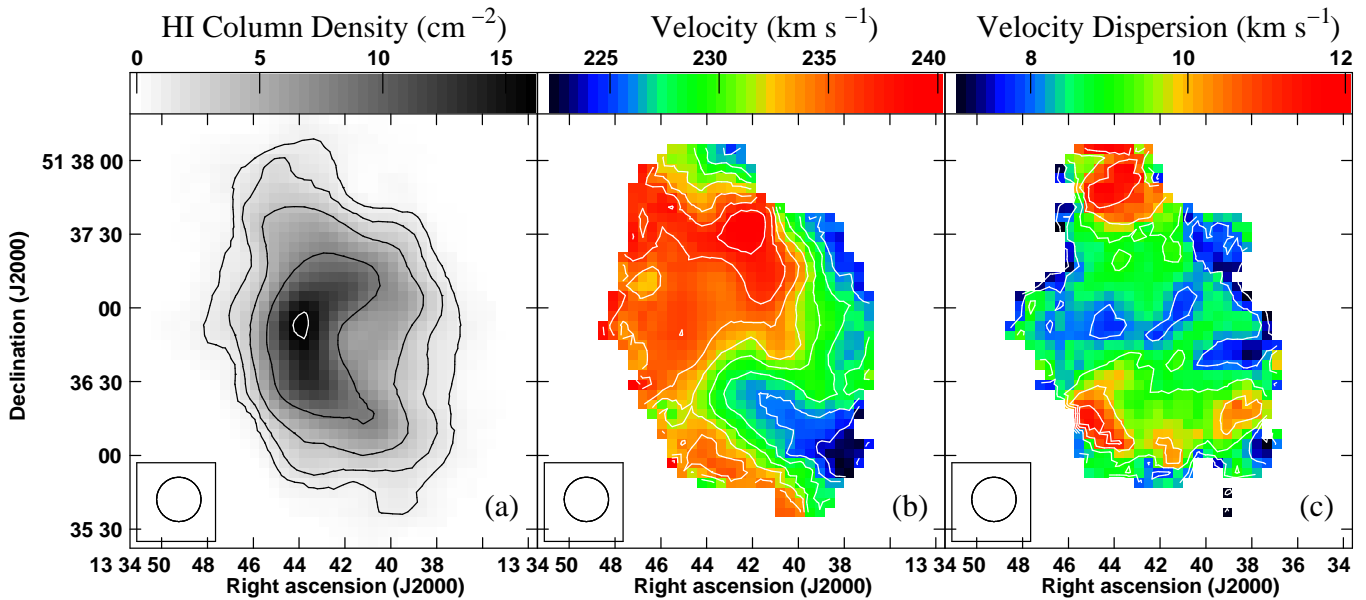}
\caption{\HI\ moment maps of NGC\,5238.  Panel (a) shows the \HI\ mass
  surface density in units of atoms cm$^{-2}$; contours are at levels
  of (1,2,4,8,16)\,$\times$\,10$^{20}$ cm$^{-2}$.  Panel (b) shows the
  intensity-weighted velocity field; contours span the range 224 --
  238 \kms, spaced by 2 \kms\ per contour.  Panel (c) shows the
  velocity dispersion of the \HI\ gas; contours span the range 7 -- 11
  \kms, spaced by 1 \kms\ per contour (note that this separation is
  formally smaller than our spectral resolution).  The
  18\arcsec\ circular beam size is shown in the bottom left of each
  panel.}
\label{figcap4}
\end{figure*}

Integrating the detected flux from NGC\,5238 produces the moment maps
shown in Figure~\ref{figcap4}.  As noted above, the high mass surface
density \HI\ gas (N$_{\rm HI}$ $\gsim$ 10$^{21}$ cm$^{-2}$, shown by
the white contour in Figure~\ref{figcap4}) displays a crescent-shaped
morphology.  The peak \HI\ column density is 1.67\,$\times$\,10$^{21}$
cm$^{-2}$, which corresponds to an \HI\ mass surface density of 13.4
\msun\,pc$^{-2}$.  We discuss the \HI\ mass surface density and its
relation to the ongoing and recent star formation in more detail in
\S~\ref{S3.2}.

The \HI\ moment one map, representing intensity-weighted
\HI\ velocity, is shown as panel (b) of Figure~\ref{figcap4}.  As
discussed in detail in \citet{mcnichols16}, the collapse of the
three-dimensional cube to two dimensions results in a more narrow
velocity width of the source; by comparing Figures~\ref{figcap2} and
\ref{figcap4}, it is clear that \HI\ gas is detected over $\sim$40
\kms\ in the data cube but that the resulting velocity field only
captures $\sim$20 \kms.  Examing the velocity field, the
\HI\ kinematics are complex.  There is a velocity gradient extending
from southwest to northeast along a position angle of roughly
20\arcdeg\ (east of north).  However, significant velocity asymmetries
are present throughout the disk; there is an S-shaped velocity profile
that may be suggestive of a warp in the disk.  We discuss the neutral
gas kinematics in detail in \S~\ref{S4}.

The \HI\ moment two map shows velocity dispersions ranging from 7-12
\kms, in good agreement with characteristic values determined from
\HI\ observations of many types of galaxies
\citep[e.g.,][]{tamburro09,ian15,mogotsi16} and in particular for
values determined for local volume dwarf galaxies
\citep[e.g.,][]{hunter2012,ott2012}.  We note that the \HI\ column density
maximum is coincident with gas with $\sigma$ $\simeq$ 8 \kms.
Similarly, in the inner regions of the disk with active star
formation, the \HI\ velocity dispersion remains below 10 \kms.

In terms of \HI\ content, NGC\,5238 can be directly compared with the
sample of 12 SHIELD galaxies presented in \citet{teich16} and
\citet{mcnichols16}. NGC\,5238 has an \HI\ mass comparable to the most
massive SHIELD galaxies; only AGC\,749237 is more \HI-massive.  The
peak \HI\ column density in NGC\,5238 is equal to the peaks found in
AGC\,110482 and in AGC\,749237.  In the following sections we further
contextualize NGC\,5238 against the SHIELD sources.

%-----------------------------------------------------------------------------%
\subsection{Comparing Stellar Population and Neutral Gas Properties}
\label{S3.2}
%-----------------------------------------------------------------------------%

NGC\,5238 is an actively star-forming dwarf galaxy (see
\S~\ref{S2.2}).  Using \halpha\ emission as a recent star formation
rate indicator (within the last 10 Myr), log(SFR$_{\rm H\alpha}$) =
$-$2.01 M$_{\odot}$\,yr$^{-1}$ .  Averaged over a longer timescale of
$\sim$100 Myr, the GALEX far-UV emission indicates a star formation rate
that is $\sim$60\% higher: log(SFR$_{\rm FUV}$) = $-$1.81
M$_{\odot}$\,yr$^{-1}$.  Each of these star formation rates are higher
than those of any of the SHIELD galaxies.

To compare the neutral gas properties with the locations of recent
star formation in NGC\,5238, in Figure~\ref{figcap5} we show
multi-wavelength imaging compared to the \HI\ mass surface density.
The fields of view are the same in all panels.  The \HI\ column
density contours highlight regions above 10$^{21}$ cm$^{-2}$.  This
mass surface density level has been empirically identified as the
requisite level for massive star formation as traced by
\halpha\ emission \citep{skillman87}; note that more recent studies
have shown that such a surface density is not universal (e.g., {Bigiel
  \etal\ 2008}\nocite{bigiel08}, {Elmegreen \& Hunter
  2015}\nocite{elmegreen15}, {Teich \etal\ 2016}\nocite{teich16}, and
the various references therein).  The aforementioned crescent-shape
morphology is very prominent in the highest column density gas.

The HST images shown in panels (a) and (b) of Figure~\ref{figcap5}
dramatically resolve the stellar populations in NGC\,5238.  The LEGUS
image \citep{calzetti15} shown in panel (a) clearly highlights the
youngest stellar clusters.  The most prominent of these structures is
located in the southern region of the galaxy (coincident with the
southern-most extension of the 1.4\,$\times$\,10$^{21}$ cm$^{-2}$
column density contour); nebular emission (arising from the
\halpha\ spectral line, which falls in the F606W filter) surrounds
this cluster in the optical image shown in panel (b).  Smaller young
clusters, identified by a comparison of panels (a) and (b), are
located in the regions of highest \HI\ column density
(N$_{\rm HI}$ $\gsim$ 16\,$\times$\,10$^{20}$ cm$^{-2}$, in the
northeast region of the optical body) as well as in the optical
center.  A close inspection of panel (b) reveals that nebular emission
is associated with all of these clusters, and is brightest near the
central, bright cluster.

As expected, the nebular emission visible in the HST image matches the
morphology of the continuum-subtracted \halpha\ image shown in panel
(d). The ground-based \halpha\ image has better surface brightness
sensitivity due to the comparatively narrow filter width; this image
clearly reveals widespread \halpha\ emission throughout the galaxy.
The most \halpha-luminous region is the central stellar cluster,
followed closely by the strong \halpha\ emission from the young
stellar cluster in the southern disk.  In addition to these two major
maxima, there is \halpha\ emission throughout the disk (including
regions associated with obvious UV clusters in the LEGUS HST image)
and also multiple dramatic loops and arcs that extend significantly
beyond the locations of the most luminous clusters (for example, the
nearly circular structure extending to the western edge of the disk).

NGC\,5238 harbors widespread UV emission in both the GALEX FUV and the
NUV bands.  As panel (c) of Figure~\ref{figcap5} shows, the UV
emission peaks in two prominent clusters that match with the highest
surface brightness \halpha\ regions and with the the prominent UV
clusters in the LEGUS images.  There is also widespread UV emission
throughout the rest of the stellar disk, including emission from
young, blue stars that trace the loop structure that is seen in
\halpha\ on the western side of the disk.

When interpreting Figure~\ref{figcap5}, it is important to note that there is
a bright foreground star in close angular proximity to the central
young cluster.  This object appears somewhat pink in the LEGUS image
in Figure~\ref{figcap5}, and is evident as an imperfectly-subtracted
image artifact in the \halpha\ image.  The SDSS spectroscopic aperture
contains this source; emission lines are nonetheless very prominent in
the resulting spectrum (not shown here), which is centered on the
\halpha\ surface brightness maximum that is slightly offset to the east.

%-----------------------------------------------------------------------------%
\subsection{Quantifying Star Formation and Neutral Gas Properties}
\label{S3.3}
%-----------------------------------------------------------------------------%

\begin{figure*}
\epsscale{1.2}
\plotone{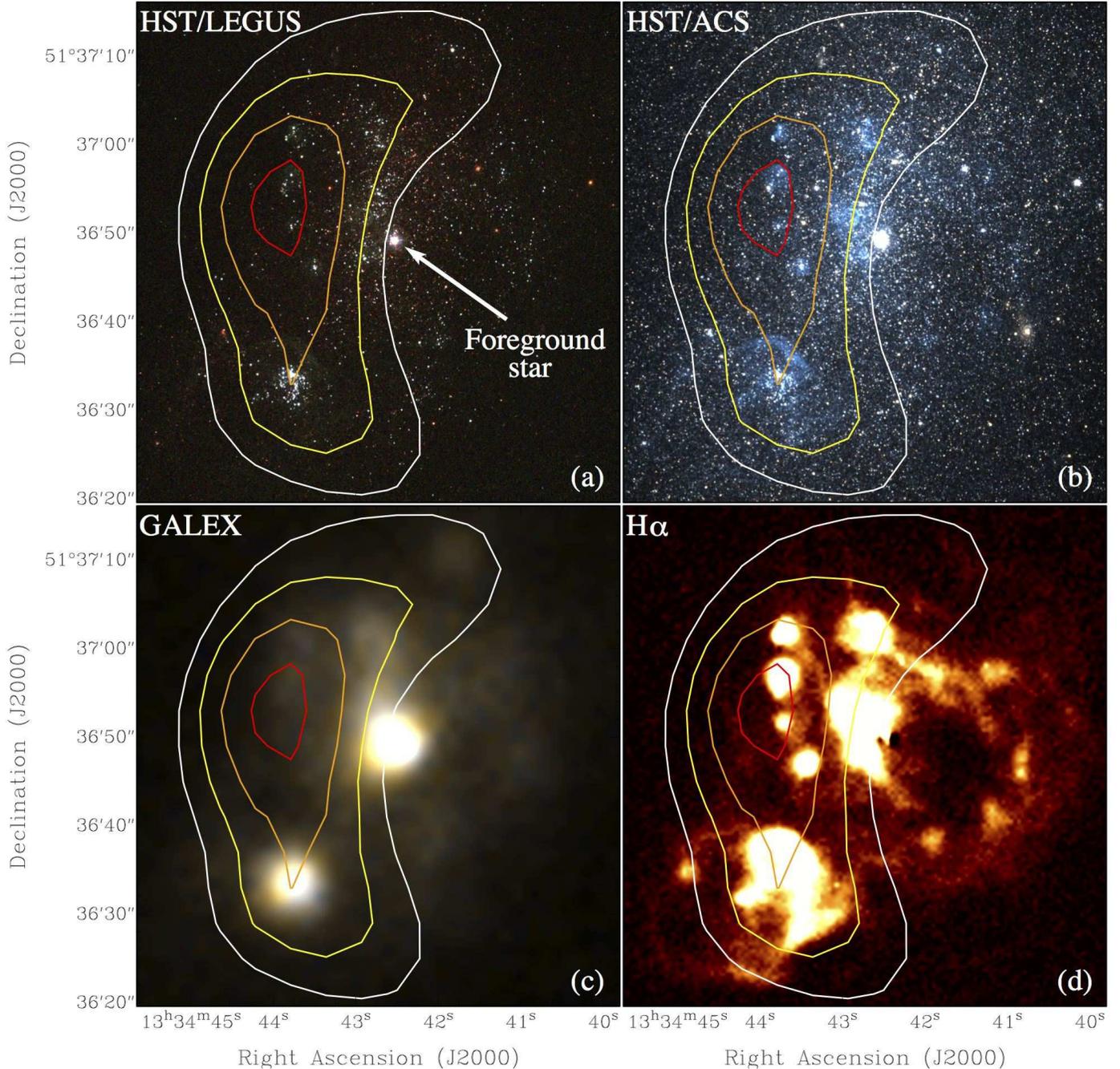}
\caption{Comparison of high mass surface density \HI\ gas with
  multi-wavelength imaging of NGC\,5238.  \HI\ column density contours
  are shown at levels of (10,12,14,16)\,$\times$\,10$^{20}$ cm$^{-2}$
  by white, yellow, orange, and red contours, respectively.  Panel (a)
  shows the 3-color HST image from the LEGUS survey (F275W in blue,
  F336W in green, F438W in red).  Panel (b) shows the 3-color HST
  image using archival ACS imaging (F606W in blue, F814W in red, and
  the linear combination of the two filters as green).  Panel (c)
  shows the UV emission detected by GALEX (far-UV channel in blue,
  near-UV channel in red, and the linear combination of the two
  filters as green).  Panel (d) shows the continuum-subtracted
  \halpha\ image from the LVL survey data products
  \citep{kennicutt08}.  The arrow in panel (a) denotes the location of
  the Milky Way foreground star.}
\label{figcap5}
\end{figure*}

As discussed in the preceding sections, the widespread recent star
formation in NGC\,5238 is occurring in a variety of physical
conditions.  The regions of highest \HI\ mass surface density are
\halpha\ and UV-luminous; however, the \HI\ column density maximum is
not coincident with the UV or with the \halpha\ maxima.  Similarly,
there is some recent star formation occurring in regions of lower
\HI\ column densities (i.e., outside of the lowest contour shown in
Figure~\ref{figcap5}, N$_{\rm HI}$ $=$ 10$^{21}$ cm$^{-2}$); such
emission is especially prominent in the western portion of the disk.  

We now seek to quantify the degree of co-spatiality (or lack thereof)
between \HI\ mass surface density and the tracers of recent star
formation (FUV and \halpha\ emission).  Numerous recent works have
investigated empirical thresholds for star formation in the neutral
and molecular ISM
\citep[e.g.,][]{skillman87,bigiel08,roychowdhury14,elmegreen15,teich16};
no single prescription has yet been identified that holds in all
environments.  For uniformity, thus we follow the procedures outlined
in \citet{teich16} by deriving the index of the ``Schmidt-Kennicutt''
relation \citep{schmidt1959}, which relates a star formation rate
surface density to a gas surface density:

\begin{equation}
\Sigma_{\rm SFR} \propto (\Sigma_{\rm gas})^{N}
\end{equation}

\noindent where the star formation rate surface density ($\Sigma_{\rm
  SFR}$) is in units of \msun\,yr$^{-1}$\,kpc$^{-2}$, the gas surface
density ($\Sigma_{\rm gas}$) in units of \msun\,pc$^{-2}$, and $N$ is
a positive number.  Specifically, we compare the star formation rate
surface densities (hereafter, $\Sigma_{\rm FUV\,SFR}$ for FUV and
$\Sigma_{\rm H\alpha\,SFR}$ for \halpha) against the \HI\ mass surface
densities (hereafter, $\Sigma_{\rm HI}$) on both a radially-averaged
and on a pixel-by-pixel basis.  Note that because of the comparatively
low star formation rates (and correspondingly low \halpha\ fluxes) of
the SHIELD galaxies, \citet{teich16} did not significantly interpret
\halpha-based radial profiles and did not show the \halpha-based
pixel-by-pixel correlation functions.  In contrast, in NGC\,5238 the
higher star formation rate and more widespread \halpha\ emission
allows us to examine trends using both instantaneous and
longer-timescale metrics.

In order to extract these diagnostics, the FUV, \halpha, and
\HI\ moment 0 images are placed on the same coordinate grid.  For the
radially averaged profiles, the HST F606W image is used to fit stellar
surface brightness as a function of radius with the
\textsc{CleanGalaxy} code \citep{hagen2014}.  This program identifies
a central aperture location (in this case, spatially coincident with
the FUV and \halpha\ surface brightness maxima), the major and minor
axis lengths of concentric ellipses in the fit, and the position angle
of those ellipses.  For NGC\,5238, the ellipses are centered at
($\alpha$,$\delta$) = 13$^{\rm h}$34$^{\rm m}$42.7$^{\rm s}$,
$+$51\arcdeg36\arcmin50.5\arcsec\ (J2000), with an axial ratio of 1.54
and a position angle (measured east of north) of 160\arcdeg.
Following the discussion in \citet{haurberg2013}, this corresponds to
an optical inclination of 50\arcdeg, which is used to deproject values
per unit area.  Surface brightnesses in \HI, FUV, and \halpha\ are
then extracted using either these concentric apertures (for the
radially-averaged method) or in individual, matched pixels (for the
pixel-by-pixel method).

\begin{figure*}
\epsscale{1.2}
\plotone{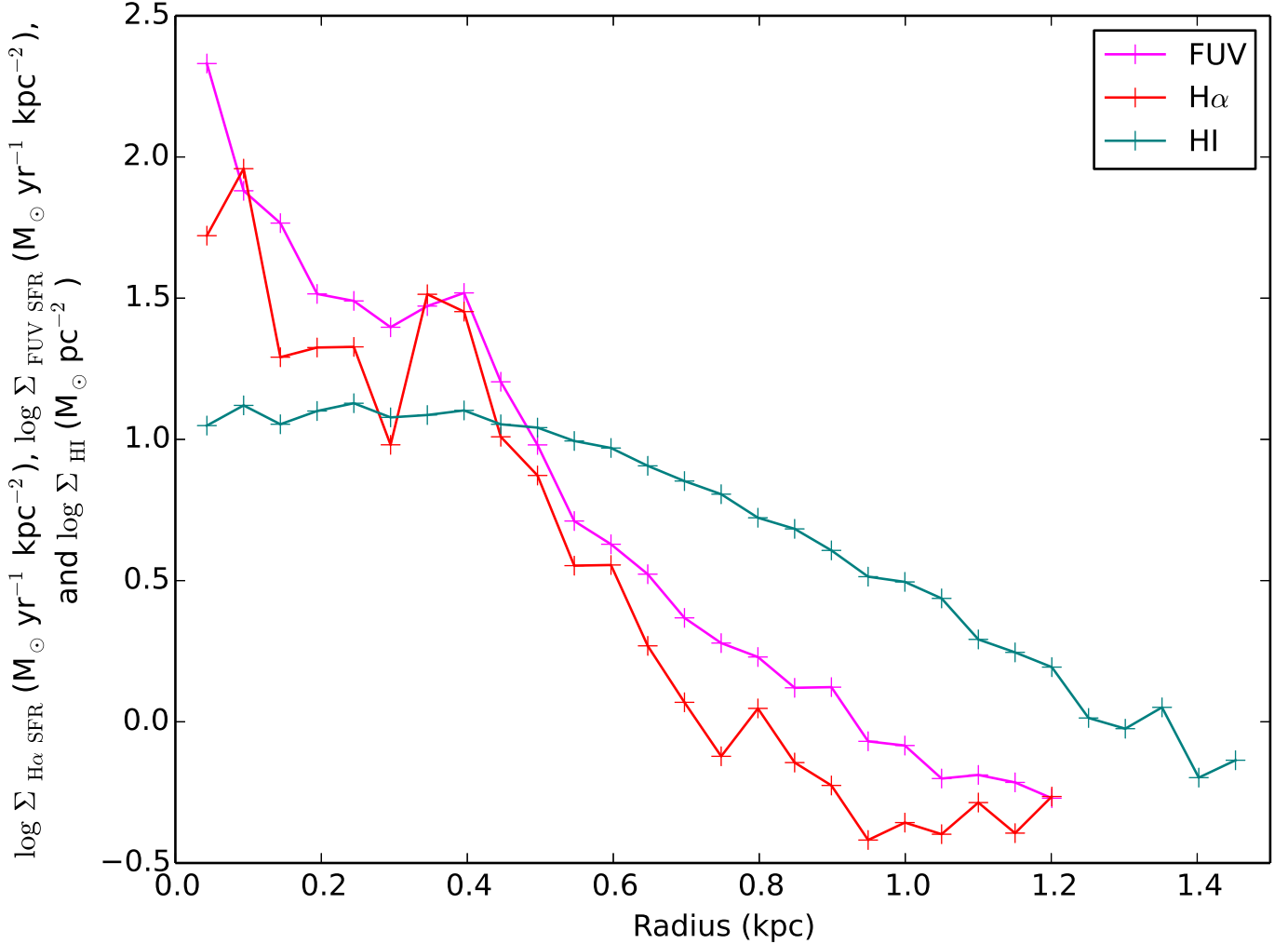}
\caption{$\Sigma_{\rm H\alpha\ SFR}$, $\Sigma_{\rm FUV\ SFR}$, and
  $\Sigma_{\rm \HI}$ and vs. radius for NGC\,5238. These plots allow us to 
examine the radially averaged trends of star formation rate surface density and 
\HI\ mass surface density.}
\label{figcap6}
\end{figure*}

The results of the radially-averaged analysis are shown in
Figure~\ref{figcap6}.  As expected based on the qualitatively similar
morphologies and the similar global star formation rates in the FUV
and in \halpha, these radial profiles are are effectively the same
using either tracer.  The \HI\ profile is less centrally peaked and
follows a shallower slope as a function of increasing galactocentric
radius.  In general, this plot demonstrates that low \HI\ mass surface
densities are typically associated with lower star formation rate
surface densities.  However, it also makes clear that the regions of
highest star formation rate surface density are associated with gas at
a range of \HI\ mass surface densities; in the inner $\sim$500 pc of
the disk, the star formation rate surface densities vary by an order
of magnitude in annuli where the \HI\ mass surface density is
effectively constant.

\begin{figure*}
\epsscale{1.2}
\plotone{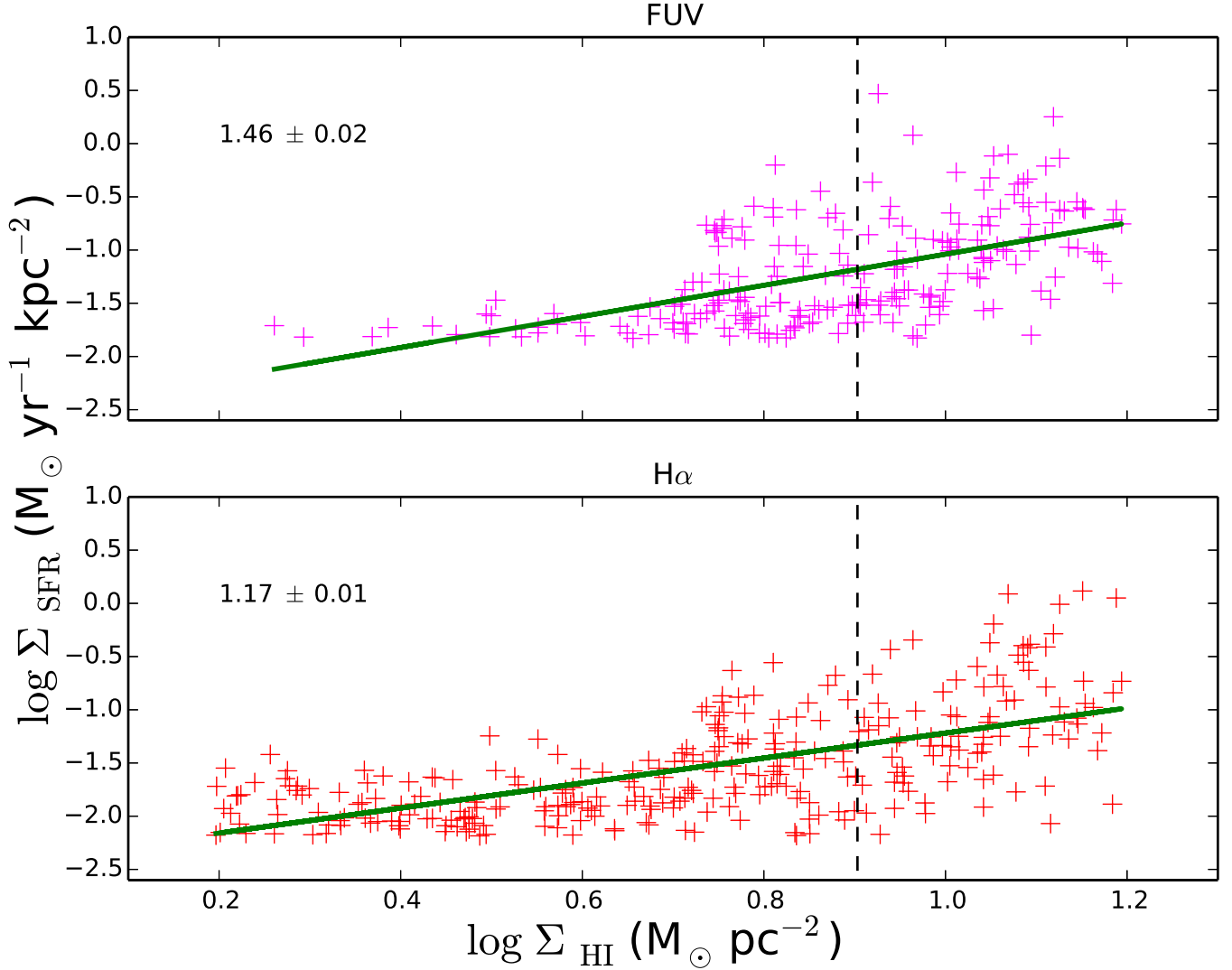}
\caption{$\Sigma_{\rm H\alpha\ SFR}$ vs. $\Sigma_{\rm \HI}$ (top) and
  $\Sigma_{\rm FUV\ SFR}$ vs. $\Sigma_{\rm \HI}$ (bottom) for
  NGC\,5238. All FUV and \halpha\ pixel values higher than 5\% of the
  maximum are plotted; for \HI\ all pixels higher than 10\% of the
  maximum are plotted.  A positive slope indicates high FUV and
  \ion{H}{1} emission in the same pixels. The dashed vertical line
  represents the column density threshold of 1$\times$10$^{21}$ atoms
  cm$^{-2}$.  The power-law best fit slopes are shown; these correspond to the 
  Kennicutt-Schmidt indices discussed in the text.}
\label{figcap7}
\end{figure*}

The radial integration used to create Figure~\ref{figcap6} smears out
localized variations in star formation rate surface density and
\HI\ mass surface density.  To overcome this limitation, we quantify
the localized relations between \HI, \halpha, and FUV emission using
the pixel-by-pixel correlation analysis shown in Figure~\ref{figcap7}.
The top and bottom panels show the FUV and the \halpha-based star
formation rate surface densities as functions of the \HI\ mass surface
density on a pixel-by-pixel basis.  These plots quantify the local
star formation law via the Schmidt-Kennicutt formalism; the
power-law index using each tracer is shown in each panel.  The FUV
slope (N $=$ 1.46$\pm$0.02) is slightly steeper than the
\halpha\ slope (N $=$ 1.17$\pm$0.01), suggesting a slightly stronger
correlation between FUV-bright regions and the highest \HI\ mass
surface densities.  However, these plots also highlight the ongoing
and recent star formation in NGC\,5238 that is occurring outside the
regions of densest \HI\ gas.  The ``canonical'' 10$^{21}$ cm$^{-2}$
column density threshold (corresponding to 7.9 M$_{\odot}$\,pc$^{-2}$)
is denoted by a vertical dotted line; while most of the highest star
formation rate surface densities occur above this \HI\ mass surface
density level, this is not a requirement.  A significant amount of
star formation is associated with \HI\ gas that is ``sub-critical''.

%-----------------------------------------------------------------------------%
\section{The Dynamics of NGC\,5238}
\label{S4}
%-----------------------------------------------------------------------------%

Low-mass galaxies such as NGC\,5238 offer a unique opportunity to
populate the low-mass, low-velocity end of the baryonic Tully-Fisher
relation ({Tully \& Fisher 1977}\nocite{tullyfisher77}; see also {McGaugh
  2012}\nocite{mcgaugh12}, {McNichols
  \etal\ 2016}\nocite{mcnichols16}, and the various references
therein).  These low-mass systems are critical for our understanding
of this fundamental scaling relation that holds over many orders of
magnitude for more massive galaxies.  Importantly, gas-rich systems in
this mass regime allow us to determine the rotational dynamics by
spectral lines; gas-poor systems must rely on dispersion measurements
from stars.

As discussed at length in \citet{mcnichols16}, extracting unambiguous
information about the dynamics of low-mass galaxies is notoriously
difficult.  That work identifies an empirical threshold of 15 \kms,
below which pressure support and coherent rotation can no longer be
differentiated.  Modeling in three spatial dimensions is preferable to
using collapsed two-dimensional representations of the data (compare,
for example, the channel maps shown in Figure~\ref{figcap2} to the
intensity weighted velocity field shown in Figure~\ref{figcap4}).
However, in most cases, degeneracies between inclination and rotation
velocity make three-dimensional work insufficiently constrained for
unambiguous solutions.

Following the methodologies in \citet{mcnichols16}, which include 
both two- and three-dimensional modeling approaches, we are not able to
produce an unambiguous model of the rotational dynamics of NGC\,5238
with the present data.  Attempts were made to model the system as an
asymmetrically distributed gas disk exhibiting coherent rotation with
a strong warp.  These challenges may be expected based on the narrow
\HI\ line width and the complex kinematics apparent in the
two-dimensional velocity field (Figure~\ref{figcap4}).

For uniformity we thus perform a spatially resolved position-velocity
(hereafter, P-V) analysis in NGC\,5238.  We first identify the
kinematic major axis by inspection as that axis along which the
projected velocity extent is largest.  This major axis P-V slice (of
width equal to the \HI\ beam size, 18\arcsec) is centered at
$\alpha$,$\delta$ = (13$^{\rm h}$34$^{\rm
  m}$42.5$^{\rm 2}$, $+$51\arcdeg36\arcmin49.0\arcsec, J2000)
and follows a position angle of
20\arcdeg\ east of north; this axis is evident in the two-dimensional
velocity field shown in Figure~\ref{figcap4}.  Note that this kinematic major
axis not aligned with the optical major axis identified in
\S~\ref{S3.3} (160\arcdeg\ east of north, or 20\arcdeg\ west of
north); the two axes differ by $\sim$40\arcdeg.  Such misalignments of
the optical and \HI\ major axes are common in the SHIELD galaxies.

\begin{figure*}
\epsscale{1.2}
\plotone{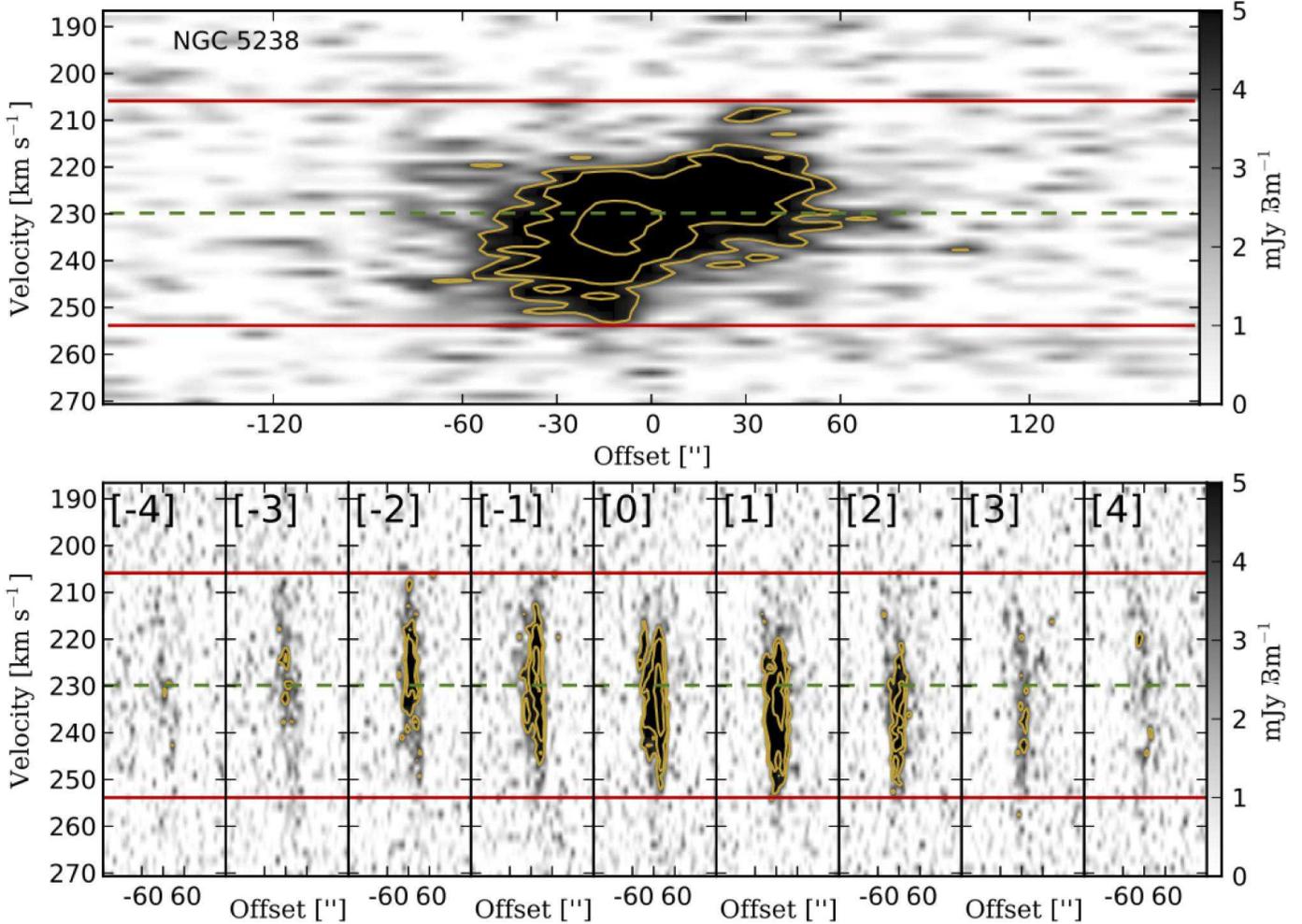}
\caption{Spatially resolved P-V analysis of NGC\,5238; all slices are
  the width of the \HI\ synthesized beam (18\arcsec). The top panel
  shows the major axis slice along a position angle 20\arcdeg\ east of
  north.  The bottom nine panels show the minor axis slices, each
  separated from one another by 18\arcsec; slice 0 passes through the
  middle of the major axis slice and through the \HI\ mass surface
  density maximum.  In all panels, the green line identifies the
  systemic velocity of the source; the red lines identify the maximal
  velocity extent to which rotation is evident; the yellow contours
  show (3, 5, 10) times the per-channel noise in the cube (1.4
  mJy\,Bm$^{-1}$).  From this analysis we infer a projected
  rotation velocity of 24 \kms, at a maximum angular extent of
  60\arcsec.}
\label{figcap8}
\end{figure*}

Minor axis P-V slices (position angle $=$ 290\arcdeg\ east of north)
are extracted at positions along the kinematic major axis, separated
by the \HI\ beam width (18\arcsec). Each minor axis slice has a width
equal to the \HI\ beam size.  In total, nine such minor axis slices
are extracted; the central minor axis slice passes through the center
of the major axis slice.  These minor axis slices show the
intensity-weighted \HI\ velocity and dispersion at a given position in
the disk.  As demonstrated in \citet{most11a}, \citet{ezbc14}, and
\citet{mcnichols16}, if projected rotation is present in a system,
then the minor-axis slices will identify any such gradient as a change
in the velocity of the \HI\ centroid as a function of position of the
slice.

In Figure~\ref{figcap8} we present the results of this spatially
resolved P-V analysis.  The major axis P-V slice identifies a total
velocity gradient of 48 \kms, spanning 60\arcsec\ (1.3 kpc at the
adopted distance of 4.51 Mpc).  This is confirmed in the minor axis
slices, which show a similar total velocity gradient across the disk.
Note that the asymmetries of the NGC\,5238 neutral gas disk result in
some turnover in the centroid velocities of the \HI\ gas in the most
distant minor axis slices (numbers $+$4 and $-$4 in
Figure~\ref{figcap8}); these slices are 72\arcsec\ (1.57 kpc) from the
slice center.  From this analysis we conclude that NGC\,5238 has a
projected rotational velocity of 24 \kms, and that this rotation can
be unambiguously identified to a maximum physical radius of 1.3 kpc.

As discussed at length in \citet{mcnichols16}, the correction of the
observed velocities for inclination is the most significant
uncertainty in the analysis of the dynamics of low-mass
galaxies. Following that work, we use the optical inclination derived
in \S~\ref{S3.3} to correct the observed \HI\ for projection effects
and thus to determine the true rotational velocity.  Our inclination
value of (50$\pm$5)\arcdeg\ compares favorably with the value of
55\arcdeg\ implied by the axial ratios derived from the UV images in
\citet{lee11}; our measurement agrees within $\sim$5\arcdeg\ with the
implied inclination of 40\arcdeg\ from the \citet{dale09} IR
apertures.

\begin{figure*}
\epsscale{1.2}
\plotone{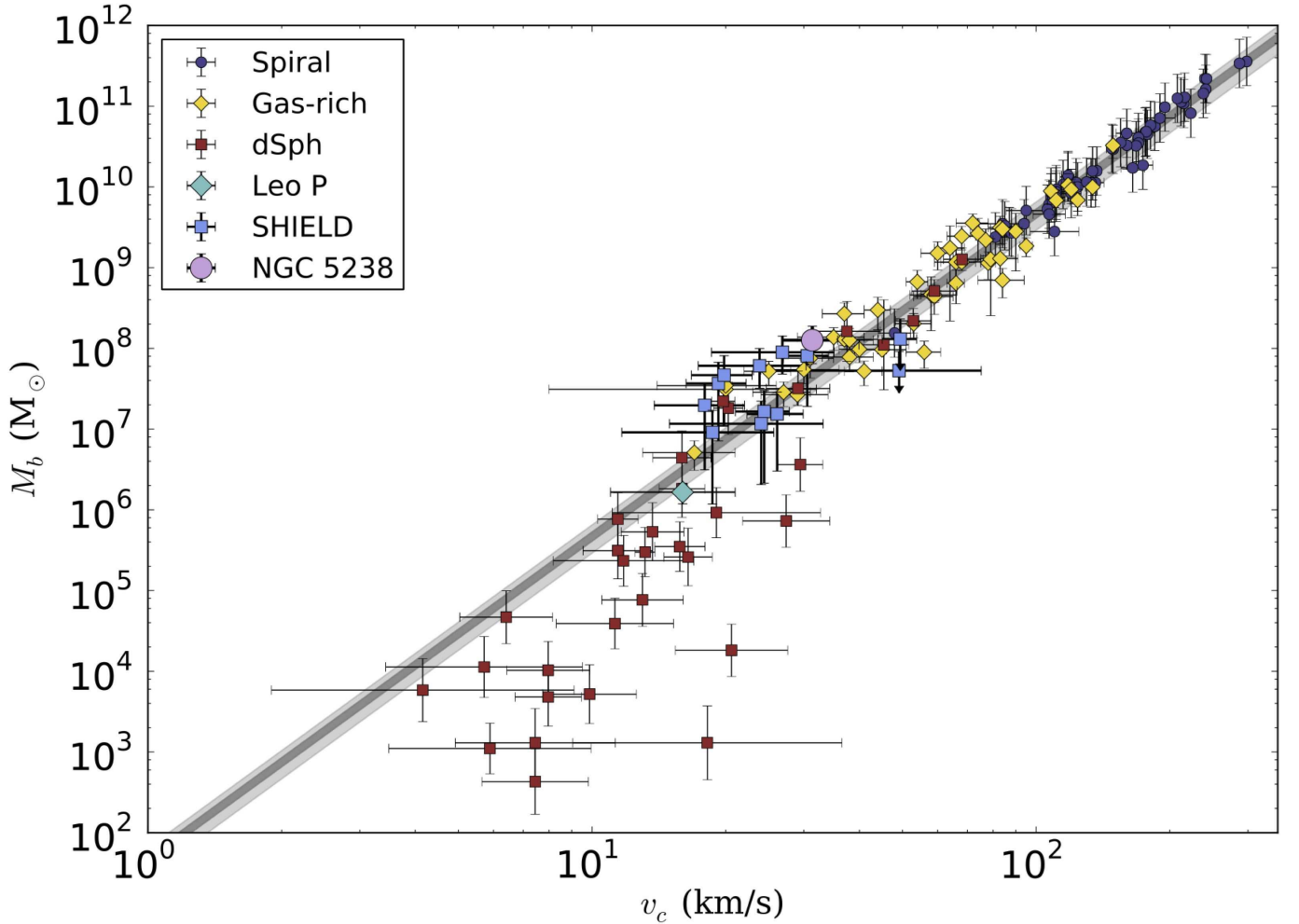}
\caption{The Baryonic Tully-Fisher relation (BTFR) as derived in
  McNichols \etal\ (2016).  The small points are drawn from McGaugh
  (2012); the purple circles correspond to spiral galaxies; the gold
  diamonds represent less massive gas-rich galaxies; the red squares
  represent spheroidal dwarf galaxies with no detectable \HI{}. The
  cyan diamond represents Leo P, the slowest rotating and lowest-mass
  galaxy known to still be relatively rich with interstellar gas
  (Bernstein-Cooper \etal\ 2014). The blue squares represent the
  SHIELD galaxies.  The large magenta circle is our new measurement
  for NGC\,5238.  The light and dark shaded gray regions represent the
  1\,$\sigma$ and the 3\,$\sigma$ deviations from a fit of the BTFR to
  the gas-rich galaxy sample, respectively.  Within measurement
  errors, NGC\,5238 lies on this calibration of the BTFR.}
\label{figcap9}
\end{figure*}

The resulting circular velocity of NGC\,5238 is 31\,$\pm$\,5 \kms,
measured to a maximum physical radius of 1.3 kpc using gas detected at
3$\sigma$ significance or higher.  The implied dynamical mass using 
\begin{math}M = \frac{v^2\,R}{G}\end{math} is
(3$\pm$1)\,$\times$\,10$^{8}$ \msun.  This can be compared to
3.7\,$\times$\,10$^{8}$ \msun\ when also explicitly accounting for
random motions in the \HI\ gas as demonstrated in \citep{hoffman96};
for this calculation we assume the average \HI\ velocity dispersion
throughout the disk (8.5 \kms; see Figure~\ref{figcap4}).  Correcting
the total \HI\ mass for Helium and other metals (a 35\% correction),
the total gas mass of NGC\,5238 is (3.7$\pm$0.3)\,$\times$\,10$^{7}$
\msun.  Using the stellar mass in Table~1, the total baryonic mass is
1.3\,$\times$\,10$^{8}$ \msun; NGC\,5238 is dark-matter dominated at a
roughly 2.4:1 ratio, a value like those found for the SHIELD galaxies
in \citet{mcnichols16}. These values allow us to place NGC\,5238 on
the baryonic Tully-Fisher relation (BTFR).  As shown in
Figure~\ref{figcap9}, NGC\,5238 agrees with the calibration of the
BTFR presented in \citet{mcgaugh12} and updated to include the SHIELD
sample in \citet{mcnichols16}.  NGC\,5238 follows the same scaling
relation between rotational velocity and baryonic mass as the SHIELD
galaxies.

%-----------------------------------------------------------------------------%
\section{Contextualizing NGC\,5238}
\label{S5}
%-----------------------------------------------------------------------------%

NGC\,5238 shares many physical characteristics with the most massive
galaxies from the SHIELD sample.  Formally, if the Right Ascension and
Declination of NGC\,5238 placed it inside the ALFALFA survey footprint
\citep{giovanelli05}, it would not have been not met all of the SHIELD
survey selection criteria.  It would meet the W$_{\rm 50}$ $<$ 65
\kms\ line width criterion but would fail the \HI\ mass criterion
(log[M$_{\rm \HI}$] $<$ 10$^{7.2}$); NGC\,5238 is $\sim$75\% too
\HI-massive to have been selected for SHIELD.  Note, however, that
stellar-based distances to many of the SHIELD galaxies
\citep{mcquinn14} moved them further than the initial distance
estimates upon selection.  The final SHIELD sample presented in
\citet{teich16} and \citet{mcnichols16} contains 4 sources with
log(M$_{\rm \HI}$) $>$ 10$^{7.2}$; one of these sources (AGC\,749237)
is more \HI-massive than NGC\,5238.

Qualitatively, the physical properties of NGC\,5238 bridge the regime
between the extremely low-mass SHIELD galaxies and the more massive
dwarf irregular galaxies in the Local Group and beyond.  As
Figure~\ref{figcap9} demonstrates, this source populates the region of
the BTFR that is sparesly populated by the highest-mass SHIELD
galaxies and the lowest-mass galaxies studied in \citet{mcgaugh12}.
The \HI\ disk of NGC\,5238 is larger (physical diameter $\simeq$4 kpc
at largest extent) than those of all of the SHIELD galaxies except for
AGC\,749237, which is of comparable size.  The peak \HI\ column
density in NGC\,5238 is the same as the highest peaks seen in the
SHIELD galaxies (AGC\,110482 and AGC\,749237).  Interestingly, the
kinematics of NGC\,5238 are significantly more confused than those of
the most massive SHIELD galaxies; AGC\,749237, AGC\,112521, and
AGC\,110482 show comparatively ordered rotation in two-dimensional
velocity fields.  NGC\,5238, in contrast, shows complex kinematics;
two- and three- dimensional rotationally-supported disk modeling fails
in this comparatively massive source.

As discussed in Section~\ref{S3.2}, the FUV and H$\alpha$ star
formation rates of NGC\,5238 are higher than those of any of the
SHIELD galaxies \citep{teich16}.  The origin of the high star
formation rate in NGC\,5238 compared to those in the SHIELD galaxies
is not immediately clear.  There are no known galaxies cataloged in
the NASA Extragalactic Database\footnote{http://ned.ipac.caltech.edu}
within a projected separation of 300 kpc whose recessional velocities
are within 100 \kms\ of the V$_{\rm HI}$ $=$ 232\,$\pm$\,1
\kms\ derived in this work (see discussions in {Lelli
  \etal\ 2014}\nocite{lelli14} and {Martinkus
  \etal\ 2015}\nocite{martinkus15}).  While the outer \HI\ disk of
NGC\,5238 does show some asymmetries, we do not detect unambiguous
evidence for an ongoing tidal interaction at the current sensitivity
and angular resolution of our VLA observations.  We note that recent
work suggests that interactions at large distances may not be
important for star formation rates in dwarf galaxies (see, e.g.,
{Pearson \etal\ 2016}\nocite{pearson16}).

The distributed recent star formation in NGC\,5238 differentiates it
from most of the SHIELD galaxies.  In those systems, almost all of the
galaxies show \halpha\ emission that is concentrated to a few isolated
HII regions or at most some faint diffuse emission.  In contrast,
NGC\,5238 harbors widespread \halpha\ emission with both clustered and
diffuse morphology.  From the far-UV perspective, the SHIELD galaxies
show a variety of morphologies; some sources have widespread far-UV
emission throughout most of the inner disk, while others show only
very faint far-UV emission.  Again NGC\,5238 stands out in a
comparative sense: the entire inner disk is UV-bright, which is unlike the 
SHIELD galaxies.  

%-----------------------------------------------------------------------------%
\section{Conclusions}
\label{S6}
%-----------------------------------------------------------------------------%

We have presented new VLA \HI\ spectral line observations of the
nearby star-forming dwarf irregular galaxy NGC\,5238.  The data
resolve the structure and dynamics of the neutral gas on spatial
($\sim$400 pc) and spectral (1.56 \kms\,ch$^{-1}$) scales that can be
meaningfully compared to other major \HI\ survey results.  The outer
portions of the neutral gas disk show significant asymmetries.  The
high mass surface density \HI\ gas displays an interesting
crescent-shaped morphology, with the highest \HI\ column densities in
excess of 1.5\,$\times$\,10$^{21}$ cm$^{-2}$.

The new \HI\ images are compared with images from HST, GALEX, and the
Bok 2.3\,m telescope in order to study the nature of the recent star
formation in NGC\,5238.  The stellar disk is resolved into multiple
stellar clusters by the HST LEGUS images \citep{calzetti15}, and there
is an exquisite match between the UV-bright stellar population and the
\halpha\ emission that permeates much of the inner disk.  The most
luminous UV clusters are associated with high column density gas
(N$_{\rm HI}$ $\simeq$ 10$^{21}$ cm$^{-2}$), but there are a
significant number of UV-bright stars in regions of lower column
densities as well.  The regions of highest \HI\ mass surface densities
are not co-spatial with the most luminous UV and \halpha\ sources.

We quantify the degree of co-spatiality between \HI\ and star
formation tracers using both radially averaged and pixel-by-pixel star
formation rate density analyses.  The far-UV results are qualitatively
similar to those found for the properties of star formation in the
SHIELD galaxies by \citet{teich16}.  The \halpha\ emission in
NGC\,5238 is significantly more widespread than in the SHIELD
galaxies. The Kennicutt-Schmidt indices are N $=$ 1.46$\pm$0.02 and N
$=$ 1.17$\pm$0.01 in the FUV and in \halpha, respectively.  The
\halpha-based value is steeper than the value found for the composite
SHIELD sample (N $=$ 0.68$\pm$0.04) in \citet{teich16}, but is in good
agreement with the values found in the individual SHIELD galaxies with
the highest star formation rates and the highest \HI\ mass surface
densities.

The complex neutral gas dynamics of NGC\,5238 preclude an unambiguous
two or three-dimensional model of the rotation of the source;
degeneracies persist between inclination and rotation velocity.  As
for the SHIELD galaxies analyzed in \citet{mcnichols16}, we use
spatially-resolved position velocity analysis to estimate the
projected rotation velocity of the galaxy.  We correct this value for
inclination based on an elliptical fit to the red stellar population
seen by HST.  The resulting rotational velocity V$_{\rm rot}$ $=$
(31\,$\pm$\,5) \kms\ implies a total dynamical mass of
3\,$\times$\,10$^{8}$ \msun.  Comparing to the sum of the neutral gas
and stellar masses, NGC\,5238 is dark matter dominated at a ratio of
roughly 2.4:1.  The galaxy falls on the baryonic Tully-Fisher relation
as presented in \citet{mcnichols16}.

NGC\,5238 is an intriguing system that warrants further study.  Higher
angular resolution \HI\ imaging could reveal small-scale feedback
processes between the multiple stellar clusters and the surrounding
neutral gas.  The stellar population is well-resolved by HST, making
possible a detailed study of the recent star formation history.  The
significant nebular emission would facilitate detailed and resolved
analysis of the chemical composition of the system.  The preliminary
estimates of the metallicity (Z $\simeq$ 19\%\,Z$_{\odot}$; {Moustakas
  \& Kennicutt 2006}\nocite{moustakas06}) place this system near the
empirical boundary below which CO emission becomes extremely difficult
to detect in dwarf galaxies
\citep[e.g.,][]{taylor98,leroy05,schruba12}.  While single-dish CO
observations by \citet{leroy05} failed to detect NGC\,5238, the high
star formation rate and significant \HI\ mass surface densities
suggest that targeted interferometric observations could be fruitful.

%-----------------------------------------------------------------------------%
\acknowledgements
%-----------------------------------------------------------------------------%

The authors are immensely grateful to the National Radio Astronomy
Observatory for making the ``Observing for University Classes''
program available to the astronomical and teaching communities.  We
would like to personally thank Vivek Dhawan, Miller Goss, Christopher
Hales, Amy Kimball, Chris Langley, Emmanuel Momjian, J{\"u}rgen Ott,
Lorant Sjouwerman, and Gustaaf van~Moorsel for making our visit to the
Domenici Science Operations Center and the VLA site productive and
enjoyable.  We are grateful to Macalester College for supporting this
project.  We would like to thank the referee for a prompt and insightful
report that helped to improve this manuscript.  

Funding for the Sloan Digital Sky Survey IV has been provided by
the Alfred P. Sloan Foundation, the U.S. Department of Energy Office of
Science, and the Participating Institutions. SDSS-IV acknowledges
support and resources from the Center for High-Performance Computing at
the University of Utah. The SDSS web site is www.sdss.org.

SDSS-IV is managed by the Astrophysical Research Consortium for the
Participating Institutions of the SDSS Collaboration including the
Brazilian Participation Group, the Carnegie Institution for Science,
Carnegie Mellon University, the Chilean Participation Group, the
French Participation Group, Harvard-Smithsonian Center for
Astrophysics, Instituto de Astrof\'isica de Canarias, The Johns
Hopkins University, Kavli Institute for the Physics and Mathematics of
the Universe (IPMU) / University of Tokyo, Lawrence Berkeley National
Laboratory, Leibniz Institut f\"ur Astrophysik Potsdam (AIP),
Max-Planck-Institut f\"ur Astronomie (MPIA Heidelberg),
Max-Planck-Institut f\"ur Astrophysik (MPA Garching),
Max-Planck-Institut f\"ur Extraterrestrische Physik (MPE), National
Astronomical Observatory of China, New Mexico State University, New
York University, University of Notre Dame, Observat\'ario Nacional /
MCTI, The Ohio State University, Pennsylvania State University,
Shanghai Astronomical Observatory, United Kingdom Participation Group,
Universidad Nacional Aut\'onoma de M\'exico, University of Arizona,
University of Colorado Boulder, University of Oxford, University of
Portsmouth, University of Utah, University of Virginia, University of
Washington, University of Wisconsin, Vanderbilt University, and Yale
University.

Facilities: \facility{VLA, HST, GALEX, Bok 2.3m}

%-----------------------------------------------------------------------------%
\bibliographystyle{apj}                                                 

%-----------------------------------------------------------------------------%

\end{document}